\documentclass[aps,twocolumn,showpacs]{revtex4}
\usepackage{amssymb}
\usepackage{graphicx}

\newcommand{\beq}{\begin{equation}}
\newcommand{\eeq}{\end{equation}}
\newcommand{\beqa}{\begin{eqnarray}}
\newcommand{\eeqa}{\end{eqnarray}}
\def\half{\frac{1}{2}}

\def\<{\langle}
\def\>{\rangle}
\def\ket#1{|#1\rangle}
\def\bra#1{\langle\, #1\,|}
\def\braket#1#2{\langle\, #1\,|\,#2\,\rangle}
\def\proj#1#2{\ket{#1}\bra{#2}}
\newcommand{\complex}{{\kern .1em {\raise .47ex\hbox {$\scriptscriptstyle |$}}\kern -.4em {\rm C}}}
\newcommand{\real}{{{\rm I} \kern -.19em {\rm R}}}

\begin{document}

\title{Trojan Horse attacks on Quantum Key Distribution systems}

\author{N. Gisin$^1$, S. Fasel$^1$, B. Kraus$^1$, H. Zbinden$^1$, G. Ribordy$^2$}

\affiliation{
    $^{1}$Group of Applied Physics, University of Geneva, 1211 Geneva 4,
    Switzerland\\
    $^{2}$id Quantique SA, 3 Ch. de la Marbrerie, 1227 Carouge/Geneva, Switzerland}

\date{\today}

\begin{abstract}
General Trojan horse attacks on quantum key distribution systems
are analyzed. We illustrate the power of such attacks with today's
technology and conclude that all system must implement active
counter-measures. In particular all systems must include an
auxiliary detector that monitors any incoming light. We show that
such counter-measures can be efficient, provided enough additional
privacy amplification is applied to the data. We present a
practical way to reduce the maximal information gain that an
adversary can gain using Trojan horse attacks. This does reduce
the security analysis of the 2-way {\it Plug-\&-Play} system to
those of the standard 1-way systems.
\end{abstract}
\maketitle

\section{Introduction}
The prominent application of quantum information science is
Quantum Key Distribution (QKD), which, together with quantum
random number generators, is the most advanced realization of
quantum devices operating at the single quanta level\cite{RMP}.
QKD offers the potential to develop for the first time in human
history provenly secure communication channels between distant
partners. The latter should be connected by a so-called quantum
communication channel, i.e. a channel able to transmit individual
quantum systems well enough isolated from the outside world such that
the receiver gets them almost unperturbed. In practice these
quantum communication channels can be realized, among others, with
standard telecom optical fibers or with free space in
line-of-sight optical channels. In both cases the transmitted
individual systems are photons. Quantum physics, in particular the
no-cloning theorem (a form of the famous Heisenberg uncertainty
relations, suitable for the analysis of QKD) guarantees that
\begin{enumerate}
\item the presence of any eavesdropper on the quantum communication channel can be
detected by the legitimate users, and
\item the legitimate users can upper bound the information that any eavesdropper could gain by
eavesdropping the quantum communication channel. Consequently, the
legitimate users can lower bound the amount of privacy
amplification they need to apply on their data in order to reduce
the eavesdropper's information to an exponentially small value.
\end{enumerate}
Accordingly, quantum physics guarantees potential\footnote{i.e.
assuming that the legitimate users do properly apply the rules of
the game, in particular that they apply enough privacy
amplification and interrupt the communication in case the detect
noise (i.e. a potential eavesdropper) is too strong to be dealt
with by privacy amplification.} security against any possible
attack on the quantum communication
channel\cite{Mayers96,BBBMR00,ShPr00,Lo01,Barbara05}.

Today a lot is known about the most powerful attacks Eve could
ever perform against the quantum channel, assuming Eve has
absolutely no technological limits, i.e. she can do everything
that quantum physics does not explicitly forbid. But, clearly,
Eve's attacks are not limited to the quantum communication
channel. For instance, Eve could attack Alice or Bob's
apparatuses, or she could exploit weaknesses in the actual
implementation of abstract QKD.

Quantum physics does not help protecting Alice and Bob's
apparatuses. Indeed, as soon as the information is encoded in a
classical physics system, it is vulnerable to copying and
broadcasting. Hence, Alice and Bob's electronics has to be
protected by classical means. Interestingly, one may ask where the
transition from quantum coding to classical coding happens. This
is an old question, the famous quantum/classical foggy transition,
but here in a modern setting: it determines what can be protected
by quantum means and what has to be protected by classical means.
But we shall not consider this question in this article. It is
anyway obvious that Alice and Bob's apparatuses need classical
protections.

Actual implementations of abstract QKD uses today's technology
(and economical constrains). Hence they necessarily move somewhat
away from the ideal scheme. It is thus of vital importance for QKD
to analyze properly the consequences of these compromises. Indeed,
some compromises might render the entire system totally insecure,
while some other compromises can be proven to maintain absolute
security, provided their analyzes are properly taken into account.
Let us stress this important point: some well implemented
compromises do not at all reduce the security of
QKD\cite{Luxx,GoLu04,TamakiLo04,Koashi05}.

An example of a very common and convenient compromise is the use
of weak laser pulses instead of the single-photon sources that are
closer to abstract qubits. This was first shown to open new
eavesdropping strategies\cite{HuIm95,BrLu00}. Next, it has been
proven that secure QKD is nevertheless possible, provided the weak
intensity of the pulses and the quantum communication channel loss
are properly taken into
account\cite{Luxx,GoLu04,TamakiLo04,Koashi05}. Finally, recently,
variations of the basic QKD protocols have been proposed that
significantly lighten the conditions for secure QKD using weak
laser pulses\cite{Wa03,ScAc04,Ko04}.

\begin{figure}
  \includegraphics[width=\columnwidth]{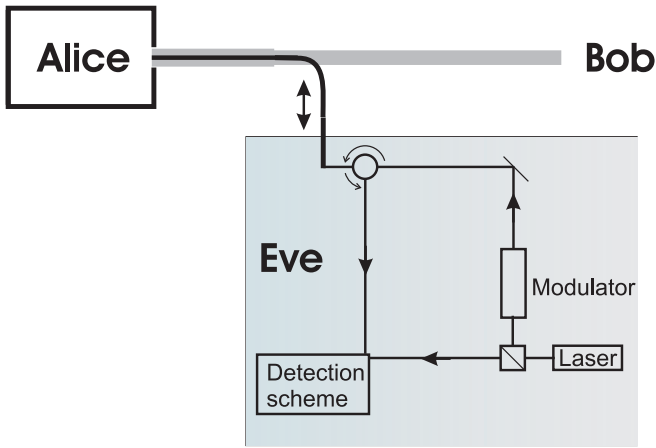}\\
  \caption{Principle of a Trojan Horse attack. Eve occupied part of the quantum
  channel (i.e. the spatial, temporal and frequency modes) to probe Alice's apparatus.
  Eve uses an auxiliary source, modulates it and analyzes the
  backscattered signal with a detector. Note that her detection scheme can rely on specificities of her
  auxiliary source, for instance on its phase.
  Eve may have to remove part of the legitimate signal, compensating the introduced loss by
  an improved quantum channel.}\label{TH}
\end{figure}

It is thus timely to study another unavoidable aspect of QKD: the
quantum channel itself is a potentially open door for an
eavesdropper into Alice and Bob's apparatuses. Indeed, even if
this door is properly designed, Eve could use it precisely at the
same time as the legitimate users: Eve could send into Alice
and/or Bob's apparatuses light pulses during the (short) times the
quantum channel is open\footnote{We consider the door as open only
during the time when it potentially gives access to some useful
information, the rest of the time the apparatus will merely
backscatters a useless signal.}, see Fig. \ref{TH}. In order to
limit this possibility, the system should be designed in such a
way that

\begin{enumerate}
\item only light at appropriate wavelength can enter (i.e. filters),
\item the "door" should be open only during short times, i.e. the encoding
 optical components should be active only during short times (i.e.
activate phase modulators only when the qubits is there), and
\item the amount of reflected light that could be exploited
by Eve is bounded by a known value.
\end{enumerate}

The purpose of this article is to analyze such attacks, known as
Trojan horse attacks. In particular we shall examine each of the
above points in section \ref{hardware}. But, first, it is useful
to get a better understanding of the techniques that such an
adversary could use, see section \ref{reflectometry}. Next, in
section \ref{securityProof} we derive the photon number statistics
of any light used in Trojan horse attacks and in \ref{theory} we
compute the maximal information that Eve could gain using Trojan
horse attacks, i.e. compute how much additional privacy
amplification is required in order to successfully combat such
attacks. Finally, in section \ref{phaseNoise} we present a simple
way to reduce this information, hence to increase the secret bit
rate.

\section{Reflectometry}\label{reflectometry}
Every optical element backscatters some amount of any incoming
light. This might be small in optical fibers (about -70dB/m) and
angle-polished connectors (typically -40dB), medium for integrated
optics components, like phase-modulators ($\approx -20$ dB) and
large for mirrors ($\ge$-1 dB).

Consequently, every optical apparatus can be examined from the
outside by shining into it well controlled light and analyzing the
backscattered light. This technique, named reflectometry, is a
standard tool for optical engineers.

For security analysis of QKD one assumes an Eve without any
technological limit. But it is useful to have an idea how the
technique works in principle and to illustrate it with today's
technology.

There are essentially two approaches to reflectometry:
\begin{enumerate}
\item Send in short optical pulses and analyze the backscattered
light intensity in function of time. From the known speed of
light, the time can be translated into distances. This technique
is called Optical Time Domain Reflectometry (OTDR), it is a very
standard tool of optical telecom engineers\cite{OTDR,OTDR2} (see
figure \ref{OTDR}).
\begin{figure}[htbp]
  \includegraphics[width=\columnwidth]{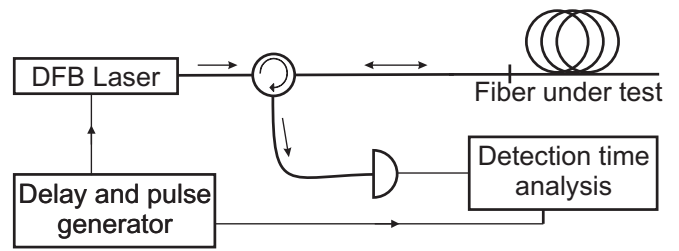}
  \caption{Functional schematic of OTDR}\label{OTDR}
\end{figure}
\item Send in coherent cw light while scanning its optical frequency
and analyze the spectrum of the backscattered light. Different
reflections correspond to different emission times, hence to
different optical frequencies. They do thus produce a beat signal.
Usually one produces on purpose one relatively large reflection
(inside the instrument) which acts as a local oscillator. The
frequency of the backscattered signal can be translated into
distance by a Fourrier transformation. This technique is called
Optical Frequency Domain Reflectometry (OFDR). It is not yet as
standard as OTDRs, but, thanks to its heterodyne detection scheme,
it holds the potential of a much larger sensitivity and dynamical
range\cite{OFDR} (see figure \ref{OFDR}).
\begin{figure}[htbp]
  \includegraphics[width=\columnwidth]{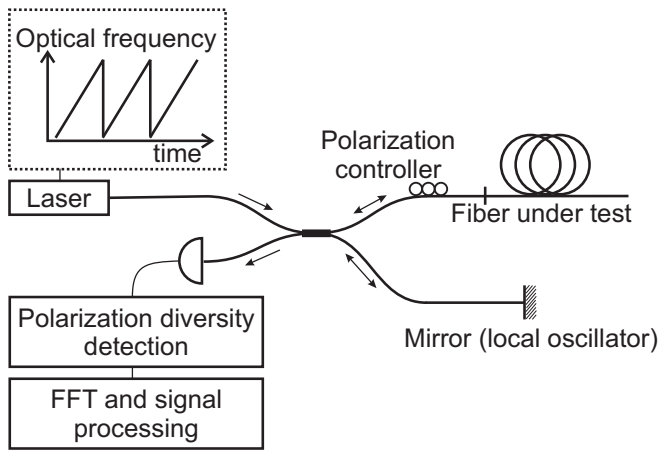}
  \caption{Functional schematic of OFDR}\label{OFDR}
\end{figure}

\end{enumerate}

The main drawback of todays OFDRs compared to OTDRs is their
limited distance range, due to the finite coherence length of the
cw laser. But, as Eve has no technological limits, we shall mainly
illustrate the potential of Trojan horse attacks using this
technique. Let us emphasize that this section is only an
illustration, clearly the counter measure by Alice and Bob should
take into account reflectometry techniques beyond today's
technique.

\begin{figure}[htbp]
  \includegraphics[width=\columnwidth]{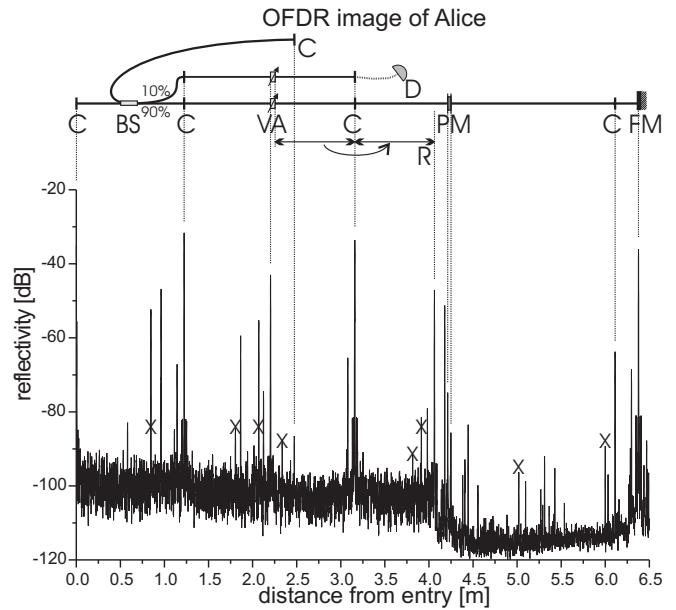}
  \caption{Example of an OFDR trace of Alice's Plug-\&-Play QKD system in which we removed the
  delay line and set the variable attenuator to its minimal value.
  A sketch of the optical circuit is displayed at the top with the corresponding reflections peaks below.
  The beam splitter (BS), connector (C), variable attenuator (VA), detector (D), phase modulator (PM) and
  Faraday mirror (FM) are all clearly visible. The peak marked R correspond to an example of multiple internal reflections. The peaks marked with a cross correspond to spurious reflections between the OFDR and Alice's components.}\label{AliceOFDR}
\end{figure}

\begin{figure}[htbp]
\begin{center}
  \includegraphics[width=\columnwidth]{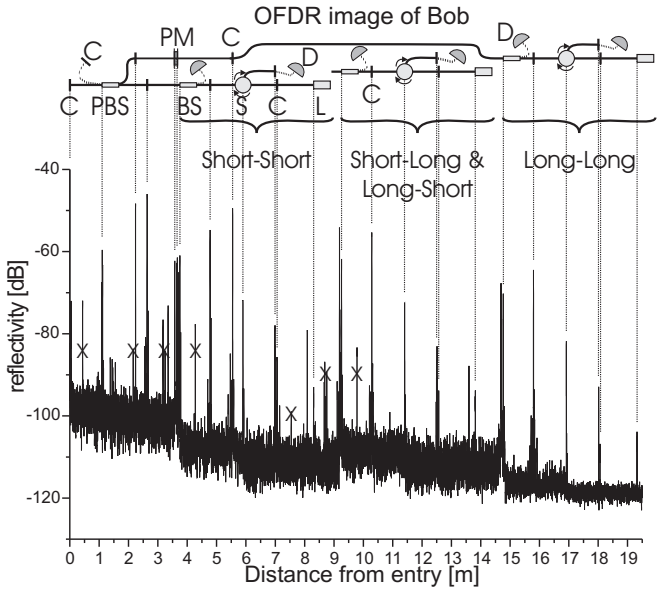}
  \caption{Example of an OFDR trace of Bob's Plug-\&-Play QKD system.
  Similar to Fig \ref{AliceOFDR}, but with the additional complication that each peak appears 3 times,
  because the incoming and reflected light both split in two, following the short and long path
  of the interferometer. For instance, one can notice that the long arm of the interferometer
  is about 11.5 meters longer than the short arm.}\label{BobOFDR}
\end{center}
\end{figure}

Fig. \ref{AliceOFDR} and \ref{BobOFDR} present the backscattered
light from Alice and Bob's apparatuses, respectively, in the case
of our Plug-\&-Play quantum cryptography
system\cite{MuGi97,RiGa98}. They illustrate that indeed quite a
lot of information can be gained by probing the apparatuses from
the outside. Let us emphasize that the same is true for all other
fiber-based apparatus, like for instance optical
amplifiers\cite{OFRDampli} and any other quantum cryptography
system. The details are given in the figure captions. Note that
for the purpose of this demonstration, we removed the about 10 km
long delay line in Alice's apparatus, because our laser (contrary
to that of Eve) has a coherence length limited to about 1 km).

\begin{figure}[htbp]
  \includegraphics[width=\columnwidth]{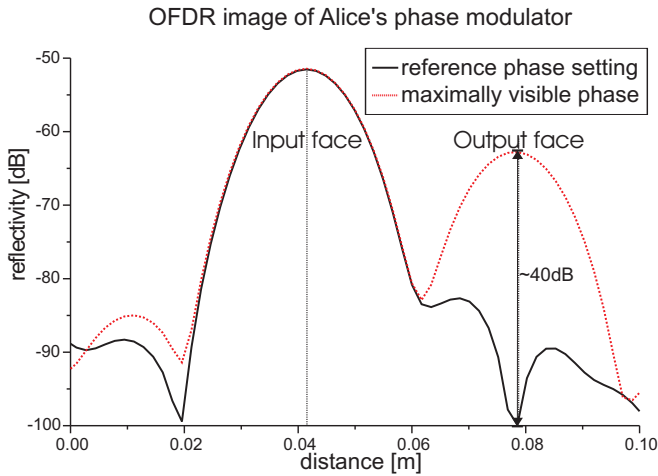}
  \caption{OFDR traces of an integrated optics phase modulator.
  Two different phase settings give raise to clearly distinguishable
  back-scatterings on the output face of the modulator. The two phase settings and the polarization of the probe light are chosen especially to exhibit a very clear effect. The measurement time is of about one second.}\label{PM}
\end{figure}

Note that it isn't yet clear how Eve could probe the setting of
the phase-modulator. However, Eve can indeed probe this setting by
exploiting the change in birefringence in Titan-indiffused LiNbO3
integrated waveguides, as illustrated in Fig. \ref{PM}. For
different kinds of phase modulators, or polarization modulators,
it is highly likely that a similar technique applies. Figure
\ref{PM} shows that it is easy to distinguish between two phase
settings of Alice's phase modulator. To obtain Fig. \ref{PM} we
had to keep the phase setting constant during about one second,
that is, a much longer time than in the usual use of the crypto
system. We also had to adjust the polarization of the probe light
and to use a polarization dependant OFDR settings, to maximize the
effect. Nevertheless, this result underlines that Trojan horse
attacks have to be analyzed seriously.

\section{Hardware counter measures}\label{hardware}
The previous section demonstrated that Trojan horse attacks on
badly designed system can be performed using today's techniques.
Consequently, every proper implementation should take care that:
\begin{enumerate}
\item the "door" lets in only wavelengths close to the operating
wavelength. Any other probe should be eliminated by properly
designed filters, and
\item the "door" should be open only during a time as short as
possible: the phase modulator, or polarization modulator, or
whatever coding device is used, should be activated only during
the short time when the legitimate signal is there.
\end{enumerate}

But even these two measures can't completely prevent Trojan horse
attacks. Indeed, Eve can multiplex her probe signal with the
legitimate signal either in polarization (if time-bin qubits are
used by Alice and Bob) or in wavelengths (Eve could reduce the
loss of the Q channel, filter out a part of the legitimate signal
and use this bandwidth for her Trojan horse attack, see fig.
\ref{TH}). Also, in practice, timing has a finite accuracy, hence
Eve can add her probes immediately before or after the legitimate
pulses.

Consequently, a first conclusion is that every sensitive apparatus
(Alice for sure, Bob depending on the protocol) must have an {\bf
active control on the intensity of the incoming light}: they
should use an auxiliary detector and monitor any incoming light.
The software should be designed such as to stop QKD as soon as
anormal intensities are detected (actually, for each qubit, there
should be a test!).

A first naive idea to circumvent the need for an auxiliary
detector is the use of attenuators and/or isolators. However,
since Eve is not limited by technology, she could merely send in
more intense light\footnote{Every physicists knows that there must
be some limit, Eve can't pulse a KJ in an ato-second pulse. At
some point, a too large energy concentration should cause the
devices to explode, melt or particle pair production starts some
nuclear reaction! But this is hard to quantify. Admittedly, the
larger the attenuator, the better.}.

A second idea could be the use of an "optical fuse", i.e. a device
that cuts the quantum channel if a to intense beam passes through
it. This is a delicate technological problem. Indeed, there is no
such fuse operating for ultra-short pulses. Hence, this does not
seem like a practical idea, though one should keep it in mind.

In practice there is a natural fluctuation in the legitimate light
and real detectors and electronics also contribute to the
fluctuation of the monitoring signal. Hence, being conservative,
one has to evaluate how much light can go to Eve without being
detected and how much information she could extract from it. Then,
appropriate privacy amplification should be applied to Alice and
Bob's data. The amount of necessary privacy amplification for any
bounded probe by Eve is computed in the next section.

\section{Statistics of Eve's probe light}\label{securityProof}
\begin{figure}[htbp]
  \includegraphics[width=\columnwidth]{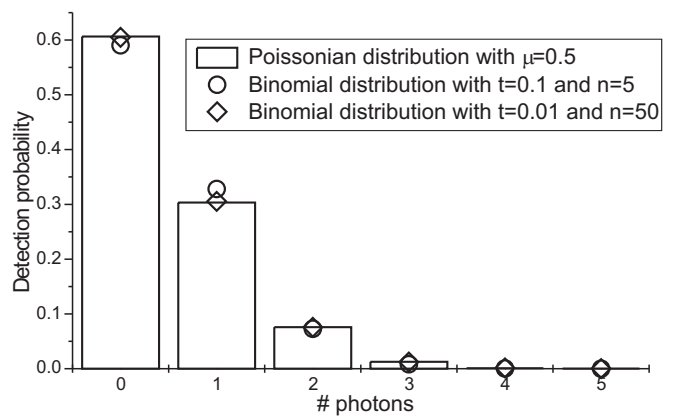}
  \caption{Comparison of photon-number distribution for poissonian and binomial distribution of the same average value.
  $\mu$: average number of photons; $t$: transmission factor for Eve's probe light, corresponding e.g the the
  attenuation at Alice's input; $n$: number of photon in the Eve's Fock-state probe light.}\label{poissvsgauss}
\end{figure}

One may question which state of light Eve should use in order to
maximize her information gain. However, it is a well known fact
that losses tend to turn any state into a state whose photon
number statistics is Poissonian. This is illustrated on Fig.
\ref{poissvsgauss} for the cases of 10 and 20 dB losses (i.e.
transmissions of 0.1 and 0.01, respectively) and mean photon
number, after attenuation, $\mu=0.5$. Since all quantum
cryptography systems (should) have attenuators and/or isolators
attenuating any light used in a Trojan Horse attack even more
severely, it is sufficient to consider light with Poissonian
statistics.

Note that this does also imply that Eve can't significantly affect
the statistics of the photon number emitted by Alice in the
Plug-\&-Play configuration, even if she replaced the intense
coherent pulse send by Bob by a squeezed state. We ellaborate on
this is section \ref{SecReduction}.

\section{Eve's potential information gain}\label{theory}
In this section we use well-known formulas to quantify the
information that Eve can extract from a weak coherent state when
she knows the "basis". Note, first that because of the huge
attenuation that any trojan horse probe light undergoes, it will
always return to Eve in a state extremely close to Poissonian, as
described in the previous section \ref{securityProof}. At best,
from Eve's point of view, it bears some coherence, that is, it is
a coherent state.

\begin{figure}
\begin{center}
  \includegraphics[width=8cm]{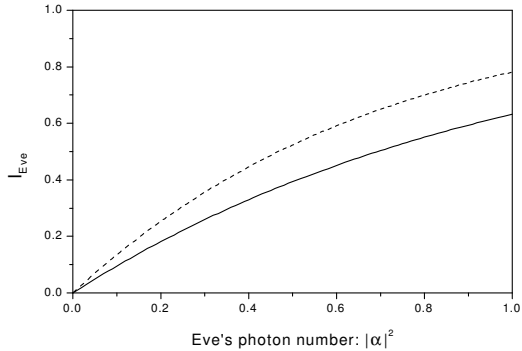}\\
  \caption{Eve's optimal information gain per qubit in function of the mean photon number
  $|\alpha|^2$ that she can collect without being detected by Alice and Bob. The upper curve
  corresponds to eq. (\ref{ITH}), the lower curve to the case that Alice and/or Bob
  applies phase randomization, eq. (\ref{Ireduced}). For example, if Alice's monitoring detector
  sets a limit to Eve's backscattered signal of 0.1 photon, then Eve may gain 0.135 and 0.095 bits
  if Alice doesn't apply or applies phase randomization, respectively.}\label{Ieve}
\end{center}
\end{figure}

Note furthermore that because of the vacuum component of the weak
coherent state, the two states corresponding to the "basis" are
not orthogonal. Explicitly, Eve has to distinguish between the
following two states $\ket{\alpha}\otimes\ket{0}$ and
$\ket{0}\otimes\ket{\alpha}$. The measurement that maximizes her
information gain is known\cite{PeresBook} and provides her with:
\beq
I_{Eve}^{Trojan}(|\alpha|^2)=1-H(p) \label{ITH}
\eeq where
\beqa
p&=&\half\big(1+\sqrt{1-|\braket{\alpha,0}{0,\alpha}|^2}\big)\label{p} \\
&=&\half\big(1+\sqrt{1-\exp{(-2|\alpha|^2)}}\big) \\
&\approx& \frac{1+\sqrt{2}|\alpha|}{2}, \eeqa

and $H$ denotes the binary entropy. Hence: \beq
I_{Eve}^{Trojan}(|\alpha|^2)\approx \frac{1}{\ln(2)}|\alpha|^2 +
O(|\alpha|^4) \eeq where $\frac{1}{\ln(2)}\approx 1.443$. This
information gain is presented graphically in Fig \ref{Ieve}.

Surprisingly, this is larger than the probability that the weak
pulse is non-empty:
\beq
Prob(non ~empty)=1-\exp{(-|\alpha|^2)}\approx |\alpha|^2
\eeq
The reason for this difference is that eq. (\ref{p}) assumes that
Eve does really hold a coherent state, i.e. that she holds a phase
reference relative to which $\alpha$ is defined. This observation
leads to a possible way to reduce Eve's maximal information gain,
as discussed in the next section.

\section{Way to reduce Eve's information}\label{phaseNoise}
Figure \ref{TH} illustrates how Eve should probe Alice and/or
Bob's apparatus in order to gain as much information about their
internal settings. Since Eve's gain can be significant, Alice and
Bob have to sacrify a significant fraction of their raw key before
obtaining a secret key. It is thus of great interest to them to
find ways to limit Eve's information. One possibility that we
present in this section, consists in Alice or Bob randomizing the
phase of $\ket{\alpha}$ relative to Eve's reference. In this way,
Eve does no longer hold $\ket{\alpha,0}$ or $\ket{0,\alpha}$,
depending on the internal setting of the apparatus, but holds the
mixed state $\rho_0$ or $\rho_1$, respectively, where:
\beqa
\rho_0&=&\int_0^{2\pi}\frac{d\theta}{2\pi}~\proj{e^{i\theta}\alpha,0}{e^{i\theta}\alpha,0}
\\
&=& \sum_{n\ge0}P\big(n|~|\alpha|^2\big)\cdot\proj{n,0}{n,0}
\eeqa
\beqa
\rho_1&=&\int_0^{2\pi}\frac{d\theta}{2\pi}~\proj{0,e^{i\theta}\alpha}{0,e^{i\theta}\alpha}
\\
&=& \sum_{n\ge0}P\big(n|~|\alpha|^2\big)\cdot\proj{0,n}{0,n}
\eeqa
where
$P\big(n|~|\alpha|^2\big)=\frac{|\alpha|^{2n}}{n!}e^{-|\alpha|^2}$
denotes the Poisson probability distribution. Eve optimal
measurement distinguishing $\rho_0$ and $\rho_1$ is also known.
Eve first measures the photon number. If she finds no photon, she
clearly gains no information. However, whenever she finds one or
more photon, then she gains full information. Hence her optimal
information gain equals the probability that the weak coherent
state $\ket{\alpha}$ is not empty:
\beq
I_{Eve}^{reduced}(|\alpha|^2)=1-P(0|~|\alpha|^2)=1-\exp{(-|\alpha|^2)}\approx
|\alpha|^2 \label{Ireduced}
\eeq

Interestingly,
$I_{Eve}^{reduced}(|\alpha|^2)<I_{Eve}^{Trojan}(|\alpha|^2)$; it
is thus of practical value for Alice and Bob to add random phases
to any light that might get back-scattered. Let us emphasize that,
clearly, these random phases act as irrelevant global phases on
the qubits, hence do not affect the proper operation of QKD, but
these random phases are relative to any possible reference that
Eve might hold, hence do reduce by the significantly factor
$\frac{1}{\ln{2}}\approx 1.44$ the maximal information that Eve
could gain using this back-scattered light\cite{LoPr05}.

\section{Reduction of security analysis of 2-way systems to 1-way systems}\label{SecReduction}
In a 2-way quantum cryptography system, like the so-called
Plug-\&-Play configuration \cite{MuGi97,RiGa98}, Eve may hold the
strong pulse that enters Alice's apparatus. Let's write
$\psi=\sum_{n\le0}c_n\ket{n}$ its state, where $\ket{n}$ denotes a
state of n photons in some appropriate mode. Note that we assume a
pure state, i.e. that the phase reference, relative to which the
complex amplitudes $c_n$ are defined, is classical. It is
straightforward to general the analysis to the case where Eve's
reference is a quantum state, i.e. Eve sends into Alice's
apparatus a state entangled with an auxiliary state held by Eve.
We like to show that phase randomization, as presented in the
previous section, together with the effect of strong attenuation
on the photon number statistics, as presented in section
\ref{securityProof}, allows one to reduce the security analysis of
2-way quantum cryptography systems to that of 1-way systems, like
those analyses in \cite{GoLu04,Luxx,TamakiLo04,Koashi05}.
Formally, phase randomization separates Eve's state $\psi$ into a
mixture of Fock number states: \beqa
\rho_{rand.ph.}&=&\int\frac{d\Phi}{2\pi}\sum_{n,m\ge0}
e^{i\Phi(n-m)}c_n c_m^* \ket{n}\bra{m} \nonumber\\
&=&\sum_{n\ge0} |c_n|^2\ket{n}\bra{n} \eeqa Next, denoting $t$ the
transmission coefficient of Alice's apparatus (go and return), one
has: \beq \rho_{rand.ph.Att.}=\sum_{m\ge0}|q_m|^2\ket{m}\bra{m}
\eeq where \beq |q_m|^2=t^m\sum_{n\ge
m}\left(\begin{array}{c}n\\m\end{array}\right)|c_n|^2(1-t)^{n-m}
\eeq Accordingly, the probability of a multi-photon pulse is: \beq
Prob(m\ge2)=\ll n(n-1)\gg\frac{t^2}{2} + O(t)^3 \eeq where
$\ll...\gg$ denote the average. For a coherent input state $\psi$,
one recovers: $Prob(m\ge2)=\frac{\ll n^2\gg
t^2}{2}=\frac{\mu^2}{2}$. For a Fock state $\psi=\ket{N}$, one
obtains, possibly surprisingly, a lower multi-photon probability:
$ Prob(m\ge2)=(N^2-N)\frac{t^2}{2}<\frac{\mu^2}{2}$.

Note again that the phase randomization separates Alice from any
possible reference-system that Eve might have prepared.
Consequently, provided Alice randomizes the global phase of each
qubit, measures the incoming intensity of each pulse and
introduces sufficient attenuation, she can bound the probability
of she sending a multi-photon pulse to Bob; hence Alice and Bob
can apply the standard security proofs to their 2-way system.

\section{Conclusion}\label{conclusion}
Trojan horse attacks should be considered for every QKD systems.
These include single-photon, weak laser pulses and continuous
variable implementations, as all necessarily include a quantum
channel that "enter" into the legitimate users apparatuses. Note
that for single-photon sources, Alice doesn't use any attenuator,
contrary to the weak pulse implementations. Hence, Trojan horse
attacks are especially dangerous for such single-photon systems.
 For the Plug-\&-Play system, the amount of reflected
light is larger than for most alternative systems. Hence, the
pressure on Eve's attacking system is reduced.

To counter such attacks, all QKD apparatuses should be properly
designed, with filters and carefully designed timing.
Additionally, auxiliary monitoring detectors must be implemented,
if not the QKD system is insecure, irrespective of the quality of
the source. Note that for the Plug-\&-Play systems, first
presented in \cite{MuGi97}, Alice does already have such an
auxiliary detector.

The accuracy of this monitoring detector determines how much
privacy amplification has to be applied in order to defeat Trojan
horse attacks. In section \ref{phaseNoise} we presented a simple
way to reduce this amount, hence to achieve larger secret keys.

\section*{Acknowledgment}
Discussions with Michele Mosca, Hoi-Kwong Lo and Norbert
L\"utkenhaus stimulated this research. This work has been
supported by EC under project SECOQC (contract n.
IST-2003-506813).


\begin{thebibliography}{99}

\bibitem{RMP} N. Gisin, G. Ribordy, W. Tittel, and H. Zbinden,
  Reviews of Modern Physics, {\bf 74}, 145 (2002).

\bibitem{Mayers96} D. Mayers, in \emph{Advances in Cryptology ---
    {CRYPTO 1996}}, LNCS {\bf 1109}, pp.~343--357, Springer (1996).

\bibitem{BBBMR00} E. Biham, M. Boyer, P.~O. Boykin, T. Mor, and V.
  Roychowdhury, in \emph{Proceedings of the 32'nd Ann.\ ACM Symposium
    on the Theory of Computing}, ACM press, pp.~715--724 (2000).

\bibitem{ShPr00} P.\ W. Shor and J. Preskill, Phys.\ Rev.\ Lett.\ {\bf
    85}, 441, (2000).

\bibitem{Lo01} H.-K. Lo, QIC {\bf 1}, 2 (2001).

\bibitem{Barbara05} R. Renner, N. Gisin and B. Kraus, quant-ph/0410215 and quant-ph/0502064.

\bibitem{Luxx} H. Inamori, N. L\"utkenhaus and D. Mayers, {\it
Unconditional security of practical QKD}, quant-ph/0107017, 2001.

\bibitem{GoLu04} D. Gottesman, H-K. Lo, N. L\"utkenhaus and J.
Preskill, Quant. Info. Comput. {\bf4}, 325, 2004.

\bibitem{TamakiLo04} K. Tamaki and H.K. Lo, quant-ph/0412035, 2004.

\bibitem{Koashi05} M. Koashi, quant-ph/0507154, 2005.

\bibitem{HuIm95} B. Huttner, N. Imoto, N. Gisin and T. Mor, Phys.
Rev. A {\bf51}, 1863, 1995.

\bibitem{BrLu00} G. Brassard, N. L\"utkenhas, T. Mor and B.C.
Sanders, Phys. Rev. Lett. {\bf85}, 1330, 2000.

\bibitem{Wa03} W.-Y. Hwang, Phys. Rev. Lett. {\bf 91}, 057901 (2003)

\bibitem{ScAc04} V. Scarani, A. Acin, G. Ribordy, and N. Gisin, Phys.
 Rev.  Lett. {\bf 92}, p. 057901 (2004).

\bibitem{Ko04} M. Koashi, Phys. Rev. Lett. {\bf93}, 120501,
2004.

\bibitem{OTDR} E-G. Neumann, {\it Single-Mode Fibers,
Fundamentals}, Ch. 13.4, Springer Series in Optical Sciences
{\bf57}, 1988.

\bibitem{OTDR2} M. Wegmuller, F. Scholder and N. Gisin, J.
Lightwave Tech. {\bf22}, 390, 2004.

\bibitem{OFDR} G. Mussi, R. Passy, J-P. Von Der Weid and N. Gisin,
J. Lightwave Techno. {\bf15}, 1-11, 1997.

\bibitem{MuGi97} A. Muller, N. Gisin, T. Herzog, B. Huttner, W.Tittel and H. Zbinden,
Applied Phys. Lett. {\bf70}, 793-795, 1997.

\bibitem{RiGa98} G. Ribordy, J.D. Gautier, N. Gisin O. Guinnard and H. Zbinden, Electron. Lett. {\bf34}, 2116-2117, 1998.

\bibitem{OFRDampli} J.P. Von Der Weid, R. Passy and N. Gisin, Photon. Tech. Lett. {\bf9}, 1253-1255, 1997.

\bibitem{PeresBook} A. Peres, ``Quantum Theory: Concepts and Methods'', Kluwer
Academic Publishers, Dordrecht (1993).

\bibitem{LoPr05} This is similar to H.-K. Lo and J. Preskill, quant-ph/0504209, though there the authors did not
consider Trojan horse attacks.

\end{thebibliography}
\end{document}